\documentclass[prd,aps,nofootinbib,preprintnumbers]{revtex4}

\usepackage{graphicx}

\begin{document}

\title{Entropy production by Q-ball decay for diluting long-lived charged particles}

\author{Shinta Kasuya$~^a$ and Fuminobu Takahashi$~^b$}

\affiliation{
$^a$ Department of Information Science,
     Kanagawa University, Kanagawa 259-1293, Japan\\
$^b$ Deutsches Elektronen-Synchrotron DESY, 22603 Hamburg, Germany}

\date{September 17, 2007}
\preprint{DESY 07-146}

\begin{abstract}
The cosmic abundance of a long-lived charged particle such as a stau
is tightly constrained by the catalyzed big bang nucleosynthesis.  One
of the ways to evade the constraints is to dilute those particles by a
huge entropy production.  We evaluate the dilution factor in a case
that non-relativistic matter dominates the energy density of the
universe and decays with large entropy production.  We find that large
Q balls can do the job, which is naturally produced in the
gauge-mediated supersymmetry breaking scenario.
\end{abstract}


\maketitle

\setcounter{footnote}{1}
\renewcommand{\thefootnote}{\fnsymbol{footnote}}

\newcommand{\bear}{\begin{array}}  \newcommand{\eear}{\end{array}}
\newcommand{\bea}{\begin{eqnarray}}  \newcommand{\eea}{\end{eqnarray}}
\newcommand{\beq}{\begin{equation}}  \newcommand{\eeq}{\end{equation}}
\newcommand{\non}{\nonumber}  \newcommand{\eqn}[1]{\beq {#1}\eeq}
\newcommand{\del}{\partial}  \newcommand{\abs}[1]{\vert{#1}\vert}
\newcommand{\vev}[1]{ \left\langle {#1} \right\rangle }
\newcommand{\ds}{\displaystyle}
\newcommand{\EV}{ {\rm eV} }
\newcommand{\KEV}{ {\rm keV} }
\newcommand{\MEV}{ {\rm MeV} }
\newcommand{\GEV}{ {\rm GeV} }
\newcommand{\TEV}{ {\rm TeV} }
\newcommand{\la}{\left\langle}
\newcommand{\ra}{\right\rangle}
\def\lrf#1#2{ \left(\frac{#1}{#2}\right)}
\def\lrfp#1#2#3{ \left(\frac{#1}{#2}\right)^{#3}}

\newcommand{\stau}{ {\tilde \tau} }

\section{Introduction}
Gauge-mediated supersymmetry (SUSY) breaking (GMSB)~\cite{GMSB} is
appealing since the problems associated with the dangerous
flavor-changing processes and CP violations are elegantly solved. In
this scenario, the gravitino mass $m_{3/2}$ is lighter than the weak
scale and therefore the gravitino is most probably the lightest
supersymmetric particle (LSP).  If the $R$-parity is conserved, the
gravitino LSP is absolutely stable and can be a good candidate for
dark matter
(DM)~\cite{Moroi:1993mb,Bolz:1998ek,Bolz:2000fu,Ellis:2003dn,Steffen:2006hw,Takahashi:2007tz}.
In addition, such a scenario with a light gravitino may lead to
spectacular collider signatures especially if the next-to-lightest
superparticle (NLSP) is a electrically charged particle such as a
stau.  The stau NLSP has a quite long life time, since the decay rate
is suppressed by the Planck scale $M_P (\simeq2.4 \times
10^{18}$\,GeV). Such longevity may enable us to measure the Planck
scale at collider experiments and therefore to test
supergravity~\cite{BHRY}, if the gravitino mass is $O(10)$ GeV.

However, the existence of such long-lived charged particles can
jeopardize the success of the big bang nucleosynthesis (BBN). Since
the stau decays during or after BBN, the energetic decay products may
alter the primordial abundance of the light
elements~\cite{Kawasaki:2004yh,Jedamzik:2004er}. Furthermore, it was
recently found that the negatively charged (long-lived) particle can
form bound states with nuclei (e.g., $^4$He$-\tilde{\tau}^-$) , which
catalyze the nuclear reactions and substantially change the abundance
of the light elements such as $^6$Li~\cite{Pospelov:2006sc}.  The
detailed discussion based on the catalyzed BBN (CBBN) limits the life
time of the stau as $\lesssim 10^3$~sec, assuming the thermal relic
abundance for the stau.  Taking the upper bound at face value, we are
led to smaller $m_{3/2}$ and/or heavier $m_{\tilde \tau}$, which are
challenging from the experimental point of view or on the basis of
naturalness.

To alleviate the BBN constraints, several solutions (e.g., $R$-parity
violation~\cite{Buchmuller:2007ui}) has been proposed. Among them, it
is simplest to assume that there is late-time entropy production to
dilute the stau abundance.  According to Ref.~\cite{HHKKY}, the
necessary dilution factor is $\Delta = (300 - 600) \times
(m_{\tilde{\tau}}/100{\rm\,GeV})$ for $m_{3/2} = 10$~GeV, where
$m_{\tilde{\tau}}$ denotes the stau mass. However, successful
late-time entropy production is not so easily achieved as one might
think of.  The reason is as follows~\cite{com-by-TY}.  To be concrete,
we assume that the gravitino is LSP while the stau is NLSP, and that a
scalar field $X$ with even $R$-parity produces large entropy. Then the
scalar mass $m_X$ must be lighter than $m_{\tilde \tau}$, since the
supersymmetric partners of the standard-model (SM) particles, if
kinematically allowed, are generically produced from the decay of
$X$~\cite{Endo:2006zj,Endo:2006ix}.  In addition, since $m_X$ is lighter than the
stau mass, $X$ must be more strongly coupled to the SM particles than
the gravitational interactions~\footnote{This ameliorates a possible
gravitino production from the scalar field $X$~\cite{Endo:2006ix}.},
in order to decay before BBN.  On the other hand, the fermionic
partner of $X$ must be heavier than $m_{\tilde \tau}$, because we have
assumed that the stau is NLSP. Therefore the mass spectrum must
satisfy $m_X < m_{\tilde{\tau}} < m_{\tilde{X}}$, where
$m_{\tilde{X}}$ is the mass of the fermionic partner of $X$.  To
realize such spectrum, however, one needs a partial cancellation
between the SUSY mass and the soft SUSY breaking mass.  Thus, it is
difficult, if not impossible, to naturally induce a late-time entropy
production in the set-up with the gravitino LSP and the stau NLSP.

In this article, we show that the successful entropy production can be
achieved by the decay of the Q ball in the minimal supersymmetric SM
(MSSM). In our scenario, a large charge $Q$ naturally makes the
effective mass smaller than $m_{\tilde \tau}$ and it also explains the
longevity of the Q ball. Since the fermionic partners are nothing but
the SM particles, there is no constraint on the fermion mass.  In
addition, a right amount of the baryon asymmetry can be generated by
the Affleck-Dine (AD) mechanism~\cite{FD,DRT}.

In the next section, we provide a brief review on the Q balls in GMSB.
We estimate the dilution factor in both cases when the Q ball dominates 
the universe after and before the freeze-out of the stau in Sec.~\ref{sec:3}.  
In Sec.~\ref{sec:4}, we show that the Q-ball decay can dilute 
the stau by the desired amount. In Sec.~\ref{sec:5} we give our conclusion and 
discussions.

\section{Q ball in GMSB}
\label{sec:2}
A Q ball is a non-topological soliton of a complex scalar field $\Phi$
given by the minimum energy configuration with a fixed $U(1)$ charge
$Q$~\cite{Coleman:1985ki}.  One of the conditions for the Q balls to
be formed is that the scalar potential with a $U(1)$ symmetry is
shallower than the quadratic potential at large field value.  The Q
balls are known to be formed associated with the scalar dynamics of
the MSSM fields, especially in connection with the AD
mechanism~\cite{FD,DRT}.  In addition, since the Q balls generically
have a very long life time, they can play important roles in
cosmology.

In MSSM, there are many flat directions composed of some combination
of squarks, sleptons and Higgs bosons. Along the flat directions, both
$F$-term and $D$-term potentials vanish in the exact SUSY limit at
renormalizable level~\cite{DRT,Gherghetta:1995dv}.  They are lifted by
the soft SUSY breaking effects, non-renormalizable operators, and
finite temperature effects. In the AD mechanism, one of the flat
directions (denoted by $\Phi$) is assumed to have a large field value
during inflation~\footnote{ Actually, multiple flat directions can
have large expectation values simultaneously, but we do not consider
this possibility for simplicity.  }. After inflation, $\Phi$ starts to
oscillate when the Hubble parameter becomes comparable to the mass of
$\Phi$. At the same time, $\Phi$ acquires the baryon (and/or lepton)
asymmetry, due to non-renormalizable baryon-(lepton-)number violating
operators that are effective only at large field values.  The scalar
potential of $\Phi$ has an approximate $U(1)$ symmetry corresponding
to the baryon and lepton symmetries that are conserved at low energy
effective theory, i.e., MSSM. As we will see below, since the scalar
potential is shallower than the quadratic potential above the
messenger scale in GMSB, $\Phi$ experiences spatial instabilities and
deforms into Q balls, where the charge $Q$ corresponds to the baryon
and/or lepton numbers.

As mentioned above, the scalar potential is lifted by the SUSY
breaking effects, non-renormalizable operators, and finite temperature
effects. In GMSB, the scalar potential above the messenger scale is
given by
\begin{equation}
\label{pot}
V(\Phi)  \simeq  M_F^4 \left( \log\frac{|\Phi|^2}{M_S^2} \right)^2  
+ c_g m_{3/2}^2 \left(1+K \log \frac{|\Phi|^2}{M_P^2}\right) |\Phi|^2
+ \frac{\lambda^2 |\Phi|^{2(n-1)}}{M_P^{2(n-3)}}
+ c_T T^4 \log \frac{|\Phi|^2}{T^2}
- c_H H^2 |\Phi|^2,
\end{equation}
where $m_{3/2}$ is the gravitino mass, $M_P \simeq 2.4 \times 10^{18}$
GeV is the reduced Planck scale, and we omit the
baryon-(lepton-)number violating operators here.  The first term comes
from the GMSB effect above the messenger scale
$M_S$~\cite{deGouvea}. $M_F$ and $M_S$ are related to the $F$ and $A$
components of a gauge-singlet chiral multiplet $S$ in the messenger
sector as
\begin{equation}
M_F^4 = \frac{g^2}{(4\pi)^4} \kappa^2 \langle F_S \rangle^2, 
\qquad M_S = \kappa \langle S \rangle,
\end{equation}
respectively, where $g$ collectively stands for the SM gauge
couplings, and $\kappa$ denotes the Yukawa coupling constant between
$S$ and the messenger fields.  In general, $M_F$ could be in the range
$10^3$ GeV $\lesssim M_F \lesssim 0.1 \sqrt{m_{3/2} M_P} \sim 5\times
10^8$ GeV for $m_{3/2} = 10$ GeV. The second term of Eq.~(\ref{pot})
comes from the gravity-mediated SUSY breaking, and the coefficient
$c_g$ is of the order unity. Here the one-loop correction is included,
and $K$ is negative with $|K|=0.1 - 0.01$ for most of the flat
directions.  Since the gravitino mass is relatively small in the gauge
mediation, this term is effective for a large field value of $\Phi$.
The third term in the potential is due to a non-renormalizable
interaction in the superpotential of the form $W_{NR} =\lambda
\Phi^n/(n M_P^{n-3})$ with $n>3$, where $\lambda$ is a coupling
constant.  The fourth term is a two-loop thermal correction to the
potential~\cite{thermal}.  The coefficient $c_T$ can be both positive
and negative depending on which flat direction we choose~\cite{KKT1},
and the absolute value is roughly given by $f_T = O(0.1)$, where we
define $|c_T| =f_T^4$.  The last term is a Hubble-induced mass term,
which stems from the quartic coupling in the K\"ahler potential
between the flat direction and the inflaton~\cite{DRT}. This term is
absent after the reheating of the inflaton.  Because of this term, the
potential has a minimum at a large field amplitude during inflation,
and the $\Phi$ field is trapped there and it serves as the initial
condition for the later dynamics~\footnote{ The same effect may be
realized by large enough Hubble-induced A terms~\cite{Aterm}.}. To be
concrete, we assume throughout this paper that the minimum is given by
the balance between the Hubble-induced mass term and the
non-renormalizable operator:
\begin{equation}
\phi_{min} \sim \left(\frac{H}{\lambda M_P}\right)^{1\over n-2} M_P,
\label{eq:min}
\end{equation}
where $\phi \equiv \sqrt{2} |\Phi|$.
The flat direction $\Phi$ traces the minimum until it starts to oscillate.

After inflation, the flat direction starts rotating, experiences
spatial instability, and deforms into Q balls~\cite{Qball1, Qball2,
KK1,KK2,KK3}.  The properties of the Q ball are well known. The charge
$Q$ is determined by the amplitude of the $\Phi$ field at the onset of
the oscillations, $\phi_{osc}$.  If the potential at $\phi=\phi_{osc}$
is dominated by the first or fourth term in Eq.(\ref{pot}), the charge
of the Q ball, which is called the gauge-mediation type, is determined
as~\cite{KK3}
\begin{equation}
\label{Qform}
Q\;=\;\beta \left(\frac{\phi_{osc}}{M(T)}\right)^4,
\end{equation}
where  $M(T)$ is defined as
\beq
M(T) = \left\{
\bear{cc}
M_F & {\rm~~for~~}M_F > f_T T_{osc} \\
f_T T_{osc} & {\rm~~for~~} f_T T_{osc} > M_F \\
\eear
\right..
\eeq
The subscript ``$osc$" denotes that the variable should be evaluated
at the onset of the oscillation of $\Phi$.  The numerical coefficient
$\beta$ is given by~\cite{KK1,KK3}
\beq
\beta \;\simeq \; \left\{
\bear{ll}
6 \times 10^{-4}\, \epsilon &{\rm~~~for~~~} \epsilon \gtrsim 0.1 \\
6 \times 10^{-5} &{\rm~~~for~~~} \epsilon \lesssim 0.1 
\eear
\right.,
\eeq
where $\epsilon \,(\leq 1)$ denotes the ratio of the baryon number
density to the number density of $\Phi$.  The size and mass of the Q
ball, and the effective mass of the field inside the Q ball (i.e., the
mass per unit charge) are written respectively as~\cite{Qball1, KK3}
\begin{equation}
R_Q \simeq \frac{1}{\sqrt{2} M(T)} Q^{1\over4}, \qquad M_Q \simeq \frac{4\sqrt{2} \pi}{3} 
M(T) Q^{3\over 4}, 
\qquad \omega_Q \simeq \sqrt{2} \pi M(T) Q^{-{1\over4}}.
\end{equation}

On the other hand, if the potential at $\phi=\phi_{osc}$ is dominated
by the second term in Eq.(\ref{pot}), the gravity-mediation type of Q
ball is formed, whose charge is given by \cite{Newtype,KK3}
\begin{equation}
Q = \beta' \left(\frac{\phi_{osc}}{m_{3/2}}\right)^2,
\end{equation}
where 
\begin{equation}
\beta' \simeq \left\{\begin{array}{lcl}
6\times 10^{-3} \epsilon & \textrm{   for   } & \epsilon \gtrsim 0.1\\
6\times 10^{-4} & \textrm{for} & \epsilon \lesssim 0.1
\end{array}\right..
\end{equation}
The size and mass of the Q ball, and the effective mass of the field
inside the Q ball (i.e., the mass per unit charge) are written
respectively as~\cite{Qball2,Newtype,KK3}
\begin{equation}
R_Q \simeq \frac{\sqrt{2}}{ |K|^{{1\over2}} m_{3/2}}, \qquad M_Q \simeq m_{3/2}Q, \qquad \omega_Q \simeq m_{3/2}.
\end{equation}

The Q ball can decay if the mass per unit charge $\omega_Q$ is larger
than the decay products that carry the same charge. For example, if
$Q$ is the baryon number, the lightest particle with baryonic charge
is a nucleon whose mass is $\simeq 1$ GeV. Therefore, such Q balls
with $\omega_Q > 1$\,GeV can decay into the nucleons (perhaps together
with other lighter particles such as $\pi$-mesons).  Since the decay
can proceed only from the surface of the Q ball, the decay rate is
bounded from above.  For the MSSM Q ball, the rate is saturated and
given by~\cite{Qdecay}
\begin{equation}
\Gamma_Q \;\simeq\; \frac{1}{Q} \frac{\omega_Q^3}{192\pi^2} 4\pi R_Q^2 \simeq 
\left\{ \begin{array}{ll} 
\ds{\frac{M_F \pi^2}{24 \sqrt{2}}Q^{-\frac{5}{4}}} & \textrm{for gauge-mediation type} \\[3mm]
\ds{\frac{m_{3/2}}{24\pi|K|} Q^{-1}} & \textrm{for gravity-mediation type} 
\end{array}\right..
\end{equation}
Therefore, the decay temperature of the Q ball is calculated as
\begin{eqnarray}
\label{TD}
T_D &\equiv& \lrfp{\pi^2 g_{*q}}{90}{-\frac{1}{4}} \left( \Gamma_Q M_P
\right)^{1\over2},\non\\ &\simeq &\left\{\begin{array}{ll}
\ds{10 \textrm{ MeV } \tilde{g}_{*q}^{-\frac{1}{4}} \lrfp{M_F}{10^7 {\rm\,GeV}}{\frac{1}{2}}
\lrfp{Q}{10^{23}}{-\frac{5}{8}}} & \textrm{for gauge-mediation type}\\[3mm] 
\ds{5 \textrm{ MeV } \tilde{g}_{*q}^{-\frac{1}{4}}  
\left(\frac{|K|}{0.01}\right)^{-\frac{1}{2}} } \lrfp{m_{3/2}}{10 {\rm GeV}}{\frac{1}{2}}
\left(\frac{Q}{10^{24}}\right)^{-\frac{1}{2}} & \textrm{for gravity-mediation type} 
\end{array}\right.,
\end{eqnarray}
where $g_{*q}$ counts the relativistic degrees of freedom at the
Q-ball decay, and we define $\tilde{g}_{*q} \equiv g_{*q}/10.75$ in
the second equality.

\section{Estimate of Dilution factor}
\label{sec:3}
Let us now estimate how much the abundance of the stau is diluted in a
situation that the Q ball releases entropy after it dominates the
energy density of the universe.  We define the dilution factor
$\Delta$ as
\beq
\frac{n_\stau}{s} = \frac{1}{\Delta} \left(\frac{n_\stau}{s}\right)_{\rm thermal},
\eeq
where the left-hand side represents the ratio of the stau number
density $n_\stau$ to the entropy density $s$ in the presence of the
entropy production due to the Q-ball decay, while the abundance of the
stau on the right-hand side is estimated without the entropy
production. Note that, although we consider the Q-ball decay, our
arguments in this section can be applied to any scenario that
non-relativisitc matter dominates the universe and decays with large
entropy production.

\begin{figure}[t]
\includegraphics[width=10cm]{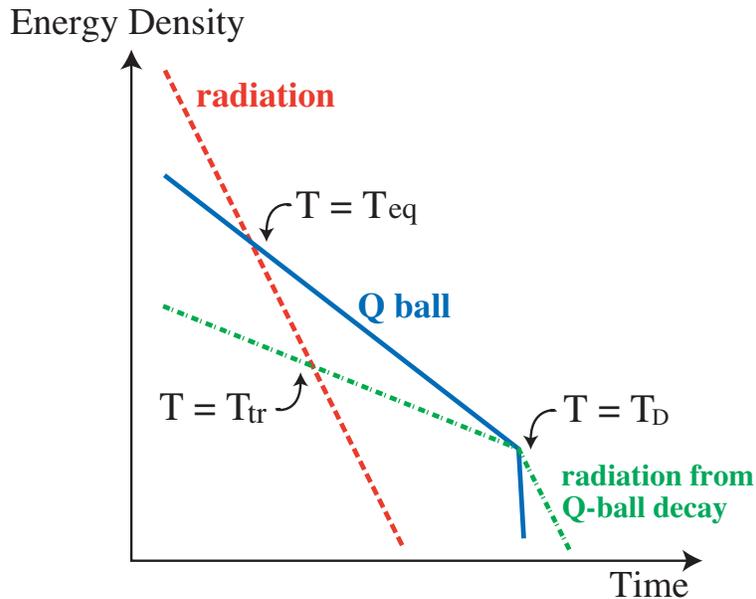}
\caption{Evolution of the energy densities of the radiation (dashed (red)), 
the Q-ball (solid (blue)),
and the radiation from the Q-ball decay (dashed dotted (green)). }
\label{fig:evolution}
\end{figure}

The stau abundance can be diluted if the Q-ball decay takes place
after the freeze-out of the stau. The dilution factor depends on the
thermal history.  We assume that the universe is radiation-dominated
before the Q balls start to dominate the energy density of the
universe.  Let us define $T_{eq}$, $T_{fo}$, and $T_D$ as the
temperatures when the Q-ball energy density becomes equal to the
radiation density, the stau freezes out, and the Q ball decays,
respectively. (We will use such notation that the subscripts $eq$,
$fo$, and $D$ denote that the variables should be estimated at $T=
T_{eq}, \,T_{fo},$ and $T_D$, respectively.)  In
Fig.~\ref{fig:evolution} we sketch the evolution of the energy
densities of the radiation, the Q ball, and the radiation produced by
the Q-ball decay.

We now consider the following cases: (i) $T_{eq} < T_{fo}$ and (ii)
$T_{eq} > T_{fo}$. In the case (i), the situation is very simple,
since the Q balls change the evolution of the universe after the
freeze-out of the stau. We thus obtain
\beq
 \left(\frac{n_\stau}{s}\right)_{\rm thermal} \;\simeq\;  \lrf{n_\stau}{s}_{eq}.
\eeq
The dilution factor is calculated as
\bea
\Delta &=& \left(\frac{n_\stau}{s} \right)^{-1}  \left(\frac{n_\stau}{s}\right)_{\rm thermal},\non\\
	    &\simeq&  \lrf{s}{\rho_Q}_D \lrf{\rho_Q}{n_\stau}_{eq} \lrf{n_\stau}{s}_{eq},\non\\
	     &\simeq& \frac{T_{eq}}{T_D},
\eea
where  $\rho_Q$ is the energy density of the Q balls.

In the latter case (ii), the stau decouples from thermal equilibrium
when the Q ball is dominating the universe. Therefore one needs to
know the dependence of the stau number density on the Hubble parameter
at the freeze-out.  Since the freeze-out takes place when the
annihilation rate becomes comparable to the expansion rate, we obtain
\beq
n_{\stau,fo} \;\sim\;\frac{H_{fo}}{\la \sigma v \ra},
\eeq
where $\langle\sigma v\rangle$ is the thermally averaged cross section
for the stau annihilation processes.  The freeze-out temperature
becomes larger than that in the usual radiation-dominated universe,
since the energy density at the freeze-out is higher in the presence
of the Q balls.  However the change in $T_{fo}$ is not significant,
because the stau decouples when it is non-relativistic, so the
freeze-out temperature is rather insensitive to the change in the
cosmic expansion rate. Thus, we simply assume $T_{fo}\simeq
m_{\tilde{\tau}}/20$ in the following~\footnote{
To include the change in $T_{fo}$, let us define $\gamma \equiv T_{fo}
/ T'_{fo}$, where $T'_{fo}$ denotes the freeze-out temperature for the
modified cosmic expansion. It is estimated as $\gamma_B \sim 1 -
(T/2m_\stau) \log(T_{eq}/T_{fo})$ for the case B, and $\gamma_C \sim 1
- (2T/m_\stau) \log(T_{fo}/T_D)$ for the case C (See Eq.(\ref{Delta0})
and Fig.~\ref{fig:evolution}). The freeze-out temperature does not
change in the case A. As long as we stick to $\Delta \sim 10^3$, the
minimum values of $\gamma_{B,C}$ are given by $\gamma_B \sim \gamma_C
\sim 0.7$, and our approximation seems to be valid.  To take account
of this change in $T_{fo}$, one has to multiply $\Delta$ in
Eq.(\ref{Delta0}) by $\gamma^{-1}$ in the cases B and C with $T_{fo}$
replaced with $T_{fo}'$. Then, one can see that $\Delta$ becomes
larger by a factor of $\sim 4$ at most. Note that including the effect
always increases the dilution factor.
}.  Such an approximation is not essential to our arguments.
Then we can estimate $\Delta$ as
\bea
\Delta  &\simeq&  \lrf{s}{\rho_Q}_D \lrf{\rho_Q}{n_\stau}_{fo} \lrf{n_\stau}{s}_{\rm thermal},\non\\
	     &\simeq& \frac{T_{fo}}{ T_D} \left(\frac{H_{fo}}{H_{fo}^{(th)}}\right),
\label{delta}	     
\eea
where we use $\rho_Q \simeq 3 H_{fo}^2 M_P^2$ at the freeze-out,
and $H_{fo}^{(th)}$ denotes the Hubble parameter at the freeze-out in
the absence of the Q-balls.

The relation betwen the Hubble parameter, $H_{fo}$, and the freeze-out
temperature, $T_{fo}$, depends on whether the newly created radiation
from the Q-ball decay dominates over the radiation that was present
from the beginning, at the freeze-out of the stau.  Let $T_{tr}$
denote the transition temperature at which both radiation components
become comparable to each other. (See Fig.~\ref{fig:evolution}.) In
the case of $T_{fo} < T_{tr}$, we have
\begin{equation}
\Delta \sim \left(\frac{T_{fo}}{T_D}\right)^3,
\end{equation}
where we use $T_{tr} \sim (T_{eq} T_D^4)^{1\over5}$ and $H_{fo} \sim
T_{fo}^4 T_D^{-2} M_P^{-1}$.  On the other hand, in the case of
$T_{fo} > T_{tr}$, we have $H_{fo} \sim (T_{fo}^3
T_{eq})^{1\over2}/M_P$. Substituting this into Eq.~(\ref{delta}), we
obtain
\begin{equation}
\Delta \sim \frac{(T_{fo}T_{eq})^{1\over2}}{T_D}.
\end{equation}

In summary, we find that the dilution factor is given as follows 
depending on when the freeze-out takes place:
\begin{equation}
\label{Delta0}
\Delta \sim \left\{ 
\begin{array}{lcl}
  \ds{\frac{T_{eq}}{T_D}} &   & ({\rm Case \ A:} \ T_{eq} < T_{fo})  \\[3mm]
  \ds{\frac{(T_{fo}T_{eq})^{1\over2}}{T_D}} &   
  & ({\rm Case \ B:} \ T_{tr} < T_{fo} < T_{eq})  \\[3mm]
  \ds{\left(\frac{T_{fo}}{T_D}\right)^3} &   
  &    ({\rm Case \ C:} \ T_D < T_{fo} < T_{tr})
\end{array}
\right..
\end{equation}
Note that the decay temperature should lie in the range of $5$ MeV
$\lesssim T_D <T_{fo}$~\cite{LowTRH} and it must satisfy $T_D <
T_{eq}$, in order to dilute the stau by the entropy production.

In the next section, we will determine the Q-ball-radiation equality
temperature $T_{eq}$ by considering the formation and the dynamics of
Q balls, in order to evaluate the dilution factor $\Delta$.

\section{Entropy production by the Q-ball decay}
\label{sec:4}
The Q-ball formation and the subsequent thermal history depend on the
scalar potential at the onset of the oscillations.  Since the large
charge Q ball is necessary for a long lifetime, the field amplitude at
the onset of the oscillation should be very large. In that case, the
zero-temperature potential would be dominated by the gravity-mediation
term (the second term in Eq.(\ref{pot})).  As we will see below, the
thermal corrections are negligible for the parameters we adopt in the
following analysis.

The flat direction starts to oscillate when $3 H \simeq m_{3/2}$. Here
we simply assume that it takes place when the inflaton oscillation
dominates the energy density of the universe.  Then the ratio of the
energy densities of the Q ball and the inflaton at the reheating is
given by
\begin{equation}
\left. \frac{\rho_Q}{\rho_{inf}}\right|_{RH} \simeq \left. \frac{\rho_Q}{\rho_{inf}}\right|_{osc} 
\simeq \frac{\frac{1}{2} m_{3/2}^2\phi_{osc}^2}{3 (m_{3/2}/3)^2 M_P^2} 
\simeq \frac{3}{2} \left(\frac{\phi_{osc}}{M_P}\right)^2.
\end{equation}
After reheating the ratio evolves as $\propto T^{-1}$, so the
Q-ball-radiation equality temperature is obtained as
\begin{equation}
T_{eq} \simeq T_{RH} \,\frac{3}{2}\left(\frac{\phi_{osc}}{M_P}\right)^2
\simeq  5 \textrm{ GeV} \left(\frac{T_{RH}}{8\times 10^7 \textrm{ GeV}}\right)
\left(\frac{\phi_{osc}}{5\times 10^{14}\textrm{ GeV}}\right)^2.
\end{equation}
Meanwhile the decay temperature is calculated from Eq.(\ref{TD}) as
\begin{equation}
T_D \simeq \left(\frac{\pi^2g_{*q}}{90}\right)^{-\frac{1}{4}} \left(
\frac{m_{3/2}M_P}{ 24\pi |K|}\right)^{1\over2}
\beta'^{-\frac{1}{2}}\frac{m_{3/2}}{\phi_{osc}} \simeq 5 \textrm{ MeV}
\left(\frac{\phi_{osc}}{5 \times 10^{14}\textrm{ GeV}}\right)^{-1}
\left(\frac{m_{3/2}}{10\textrm{ GeV}}\right)^{3\over2},
\label{eq:td}
\end{equation}
where $|K| = 0.01$ and $g_{*q} = 10.75$ are used.  Since the
freeeze-out temperature is $T_{fo} \sim 5$ GeV for $m_\stau = 100$
GeV, the dilution factor is estimated as in the case A:
\begin{equation}
\Delta \sim \frac{T_{eq}}{T_D} \sim 10^3
\left(\frac{T_{RH}}{8\times 10^7 \textrm{ GeV}}\right)
\left(\frac{\phi_{osc}}{ 5 \times 10^{14}\textrm{ GeV}}\right)^3
\left(\frac{m_{3/2}}{10\textrm{ GeV}}\right)^{-{3\over2}}.
\end{equation}
In this case we have $Q\sim 10^{24}$.  Notice that $T_D$ shown in
(\ref{eq:td}) can marginally satisfy the BBN constraints
\cite{LowTRH}. For slightly larger $m_{3/2}$ or smaller $\phi_{osc}$,
one can have a large enough $T_D$ that safely satisfies the BBN bound,
keeping the dilution factor $\Delta$ large enough.  Since the initial
amplitude of the flat direction is determined as in Eq.(\ref{eq:min}),
it will be realized for the $n=6$ direction ($LLe$ or $udd$) with
$\lambda\simeq 0.006$, or the $n=7$ direction ($dddLL$) with
$\lambda\simeq30$.  The coefficient of the thermal logarithmic
corrections will be negative for these directions. If the thermal
correction to the potential dominates over the gravity-mediation term,
it will spoil the above scenario because the $\Phi$ field will be
trapped by the negative thermal logarithmic potential, and so, it
cannot be released for a long time.  In order to avoid such a
situation, we must impose a condition that the thermal logarithmic
correction is negligible at the onset of the oscillations: $f_T^4
T_{osc}^4 < \frac{1}{2} m_{3/2}^2\phi_{osc}^2$~\footnote{
The thermal logarithmic term may not appear in such situation that the
dilute plasma before reheating is suppressed because of e.g., the
small mass of the inflaton and/or the existence of appropriate
multiple flat directions.
}. 
Using $T_{osc}^4 \simeq 0.5 \,H_{osc} M_P T_{RH}^2$, it is rewritten as
\begin{equation}
\phi_{osc} > \frac{f_T^2 T_{RH}}{\sqrt{3}} \left(\frac{M_P}{m_{3/2}}\right)^{1\over2}
\sim 2\times 10^{14} {\rm\, GeV}
\lrfp{f_T}{0.1}{2}
 \left(\frac{T_{RH}}{8\times 10^7\textrm{ GeV}}\right)
 \lrfp{m_{3/2}}{10{\rm GeV}}{\frac{1}{2}},
\label{eq:const} 
\end{equation}
which is satisfied in the above analysis.

Finally, we comment on the last moment of the Q-ball decay.  As the
charge becomes small, the gravity-mediation type Q ball gradually
deforms into the gauge-mediation type one.  Therefore, at a certain
point, the mass per unit charge of the Q ball may exceed the stau mass,
$\omega_Q > m_\stau$, which implies that the stau
can be produced.  In order to suppress the stau abundance as $Y_\stau
\lesssim 10^{-17}$, we must impose $M_F\lesssim 10^5$ GeV.  This is
because $Y_\stau \sim B_\stau (T_D/m_{3/2} )$ and the branching ratio
is estimated as $B_\stau \sim Q_{cr}/Q$ where $Q_{cr} \sim
(M_F/m_\stau)^4$. Such a small value of $M_F$ is realized in a model
where the Yukawa coupling $\kappa$ is suppressed as in the case of the
composite $S$ field.

\section{Conclusion}
\label{sec:5}
We have shown that the Q-ball decay can produce large enough entropy
to dilute the cosmic abundance of a long-lived charged particle such
as a stau.  Since the Q balls are composed of the MSSM particles, our
scenario is minimal in some sense.  Successful late-time entropy
production is not so easily achieved as one might think of: the
particle that produces entropy should have a smaller mass than the
stau and a long lifetime, and no unwanted particle production should
occur.  In our scenario, a large charge $Q$ naturally makes the
effective mass smaller than the stau mass, and the decay products are
just the SM particles. It also explains the longevity of the Q
ball. The large Q ball can be naturally produced in the dynamics of
the flat direction.

In addition, we have derived analytically the dilution factor for the
cases that the stau freezes out both before and after the Q ball
starts to dominate the universe.

Lastly let us briefly discuss the baryon asymmetry and the dark matter.
Due to the late-time entropy production, the baryon asymmetry and the
gravitino dark matter are also diluted.  One way out is to
over-produce both by the amount of the dilution beforehand. Another is
to create them after the entropy production.  As for the gravitino
dark matter, the thermal production does not suffice since the
reheating temperature cannot be too high due to the constraint
(\ref{eq:const}).  Therefore, one has to rely on the non-thermal
production~\cite{Takahashi:2007tz,Kawasaki:2006gs}.  On the other
hand, one can obtain a right magnitude of the baryon asymmetry from
the Q-ball decay itself, making use of the Affleck-Dine baryogensis.
If the flat direction has the baryon number, it must start the oscillations 
with suppressed angular motion (i.e., $\epsilon \sim O(10^{-7})$). 
On the other hand, for the leptonic direction, the lepton 
charges evaporated from the Q ball before the electroweak phase transition 
are partially converted into the baryon asymmetry through the sphaleron 
processes.  In this case, $\epsilon \sim O(10^{-3})$ is necessary to have a 
right abundance of the baryon asymmetry.

\section*{Acknowledgments}
SK is grateful to M. Kawasaki for useful discussion.  FT thanks M. Endo for
discussion. The work of SK
is supported by the Grant-in-Aid for Scientific Research from the
Ministry of Education, Science, Sports, and Culture of Japan,
No.~17740156.



\end{document}